\documentclass[modern]{aastex631}

\newcommand\msun{M_{\odot}}
\newcommand\mhi{M_{HI}}
\newcommand\vsys{V_{sys}}
\newcommand\wfty{W_{50}}

\newcommand{\kms}{{\ensuremath{\mathrm{km\,s^{-1}}}}}

\defcitealias{2023Zhou}{Z23}

\begin{document}
\newcommand\hi{\ion{H}{1}}

\title{A Bigger Cloud 9? New HI Observations of the RELHIC Candidate M94-Cloud 9}

\correspondingauthor{Ananthan Karunakaran}
\email{ananthan.karunakaran@utoronto.ca}

\author[0000-0001-8855-3635]{Ananthan Karunakaran}
\affiliation{Department of Astronomy \& Astrophysics, University of Toronto, Toronto, ON M5S 3H4, Canada}
\affiliation{Dunlap Institute for Astronomy and Astrophysics, University of Toronto, Toronto ON, M5S 3H4, Canada}

\author[0000-0002-0956-7949]{Kristine Spekkens}
\affiliation{Department of Physics and Space Science, Royal Military College of Canada P.O. Box 17000, Station Forces Kingston, ON K7K 7B4, Canada}
\affiliation{Department of Physics, Engineering Physics and Astronomy, Queen’s University, Kingston, ON K7L 3N6, Canada}

\begin{abstract}
We present new \hi{} observations of the REionization-Limited \hi{} Cloud (RELHIC) candidate, M94-CL9, detected around M94 by Zhou et al.\ using the Five-hundred-meter Aperture Spherical Telescope (FAST).\ M94-CL9's \hi{} properties as detected by FAST are consistent with a RELHIC as noted by Benitez-Llambay \& Navarro.\ Our observations with the Robert C.\ Byrd Green Bank Telescope (GBT) detect greater \hi{} emission in M94-CL9 and result in \hi{} properties that are larger (corrected velocity width, $W_{50,c,t}=35.7\pm0.6\,\kms$; and integrated flux, $\int\mathrm{Sdv}=0.28\pm0.04\,\mathrm{Jy}\cdot\kms$) than those found by Zhou et al.\ but that match those from the FAST All-Sky \hi{} (FASHI) survey.\ These larger properties do not preclude M94-CL9 from being a RELHIC, but the wider spectral extent and spectral asymmetry reported here may be in tension with predictions of RELHIC properties.\

\end{abstract}


\section{Introduction} \label{sec:intro}
Within the $\Lambda$CDM paradigm, low-mass dark matter haloes and the galaxies that are embedded within them are the building blocks of large-scale structures in the Universe.\ Some low-mass haloes ($M_{200}\sim10^9\msun$), despite having significant \hi{} reservoirs, may have had their potential for significant star formation halted by reionization.\ Using the APOSTLE cosmological simulations, \citet{2017BenitezLlambay} postulate that there exist large quantities of low-halo mass ($3\times10^8 < M_{200}/\msun < 5\times10^9$) systems with significant \hi{} reservoirs that correlate well with their halo masses yet have minimal stellar components.\ In simulations, the \hi{} components of these REionization-Limited \hi{} Clouds (RELHICs) are generally circular and symmetric in their distributions, have low velocity widths ($\wfty\sim20\,\kms$), are well-separated from other massive haloes, and are in approximate hydrostatic and thermal equilibrium with their dark matter halo and the ionizing UV background, respectively.\ The implications of discovering and characterizing such objects are significant for $\Lambda$CDM.

Using the Five-hundred-meter Aperture Spherical radio Telescope (FAST), \citet[][hereafter \citetalias{2023Zhou}]{2023Zhou} observed the environment of the Messier 94 (M94) galaxy, revealing extended \hi{} emission, including three new \hi{} clouds.\ One of these clouds, M94-Cloud 9 (M94-CL9), has no obvious optical counterpart and its \hi{ properties } from FAST are comparable to those of RELHICs in the APOSTLE simulations: it has a narrow velocity width ($\wfty\sim20\kms$), it is relatively isolated from M94 ($D_{proj}\sim109\,\mathrm{kpc}\sim1.3$ degrees), and its \hi{} distribution is fairly circular ($\theta_{CL9} \sim 6'$, axial ratio $\mathrm{b/a}>0.8$ at a column density of $1\times10^{18}\mathrm{cm^{-2}}$; $\theta_{FAST}\sim2.9'$).\ More recently, \citet{2023BenitezLlambay} presented a comparison of the FAST \hi{} distribution of M94-CL9 measured by \citetalias{2023Zhou} with RELHIC models from \citet{2017BenitezLlambay} and find consistency between observations and simulations.\

Here, we present a new \hi{} profile of M94-CL9 using the Robert C.\ Byrd Green Bank Telescope (GBT).\ The larger, 9' {\it L}-band beam of the GBT relative to the 2.9' one for FAST should encompass the entire FAST detection presented by \citetalias{2023Zhou} and provides a useful comparison spectrum.

\section{Observations and Results} \label{sec:style}
We conducted 42 minutes of position-switched observations of M94-CL9 with the GBT (program AGBT23B-099).\ Along with the \textit{L-}band receiver, we used a narrow bandpass configuration (bandpass width $\sim$ 11.72 MHz/2500$\,\kms$, spectral resolution $\sim$ 0.36 kHz/0.7$\,\kms$) of the Versatile GBT Astronomical Spectrometer (VEGAS).\ We reduced the data using the standard \textit{getps} procedure within GBTIDL\footnote{http://gbtidl.nrao.edu/}.\

\begin{figure}
    \centering
    \includegraphics[width=\textwidth]{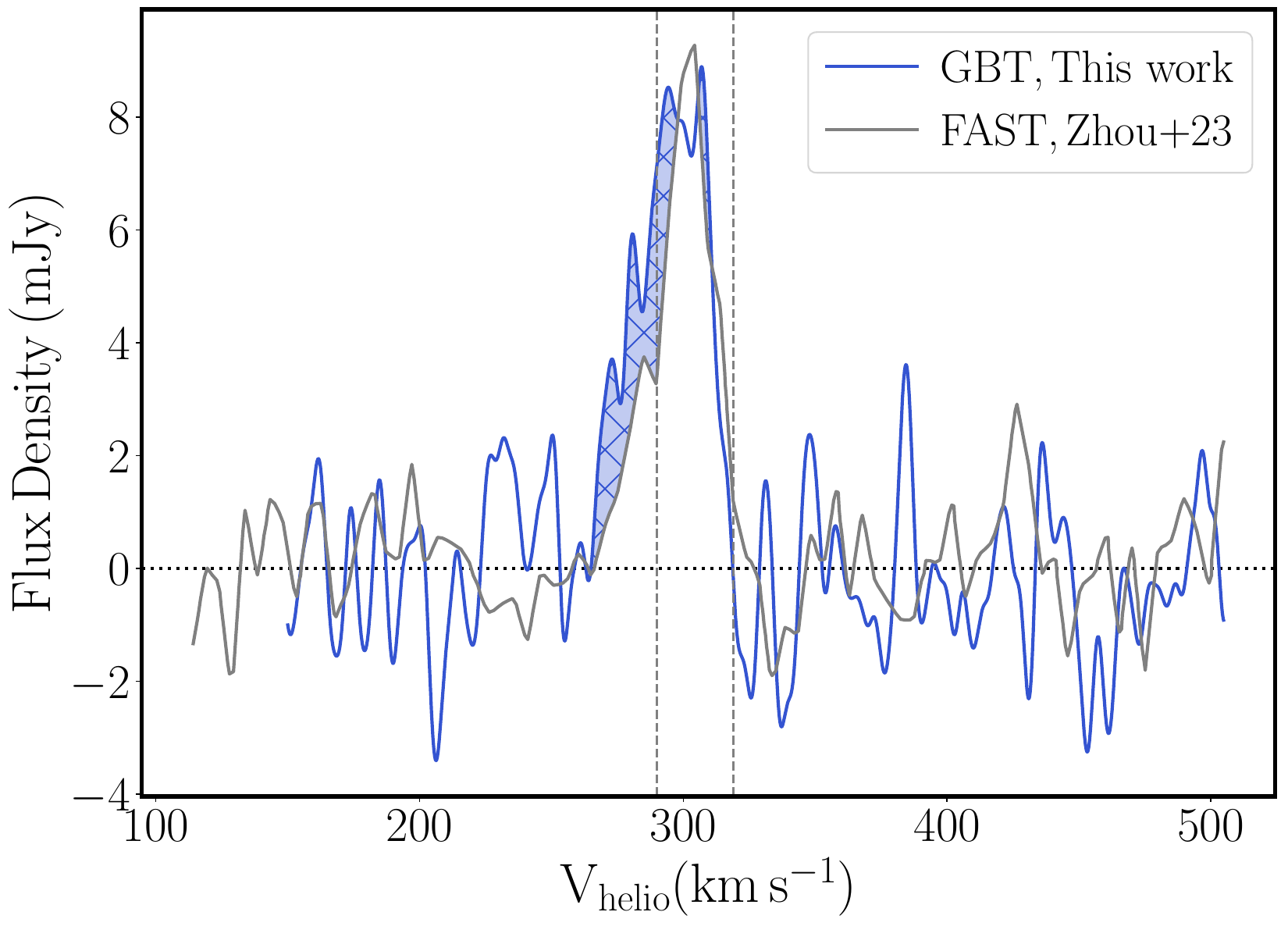}
    \caption{GBT \hi{} spectrum of M94-CL9 shown in blue and the FAST spectrum from \citetalias{2023Zhou} shown in gray.\ The velocity range of the \hi{} emission noted by \citetalias{2023Zhou} is shown with the vertical dashed lines.\ We highlight the difference in \hi{} emission as the hatched and shaded regions.\ }
    \label{fig:GBTspectrum}
\end{figure}

The calibrated GBT spectrum\footnote{We note that we have scaled the flux output by \textit{getps} by a factor of 1.2 following the results of \citet{2020Goddy} to account for offsets in noise diode calibration values.} is shown in Figure \ref{fig:GBTspectrum} in blue, and has been smoothed to a spectral resolution of $4.88\,\kms$ to match that of the FAST observations from \citetalias{2023Zhou} in gray.\ The representative noise of the GBT spectrum at this resolution is $\sim1.2$ mJy.\ We show the velocity range over which \citetalias{2023Zhou} consider the \hi{} emission of M94-CL9 in the FAST data (see their table 1) with the vertical gray lines.\ 

We derive the distance-independent properties (i.e.\ systemic velocity, $\vsys$; uncorrected velocity width, $\wfty$; integrated flux, $\int\mathrm{Sdv}$) of the \hi{} emission following \citet{2020bKarunakaran}.\ We find $\vsys=293.8\pm0.4\,\kms$, $\wfty=37.6\,\kms$, and $\int\mathrm{Sdv}=0.28\pm0.04\,\mathrm{Jy}\cdot\kms$.\ The velocity width after applying corrections for instrumental broadening, redshift broadening, and turbulence is $W_{50,c,t}=35.7\pm0.6\,\kms$.\ The resulting \hi{} mass, $\mhi$, derived using the standard equation \citep[e.g.][]{1984Haynes} and assuming the M94 distance of 4.66 Mpc\footnote{We believe they have assumed a distance uncertainty of 0.5 Mpc and for consistency adopt this as well.} as in \citetalias{2023Zhou}, is $\mhi=(1.4\pm0.4)\times10^6\msun$.\

\section{Discussion} \label{sec:floats}
From Figure \ref{fig:GBTspectrum} we can see that the \hi{} profiles obtained with FAST\footnote{We note that the spectra shown in figure 4 of \citetalias{2023Zhou} should have units of flux density (as plotted here) instead of antenna temperature since the peak fluxes not only match here but also match the listed peak flux value ($9.03\,\mathrm{mJy}\cdot\kms$) in the FASHI survey catalog.} (gray) and the GBT (blue) are consistent with one another.\ Our integrated flux ($0.28\pm0.04\,\mathrm{Jy}\cdot\kms$) is comparable to that listed in table 1 of \citetalias{2023Zhou}, $0.239\pm0.01\,\mathrm{Jy}\cdot\kms$, but a factor of two larger than the value listed in the text of their section $4.2$, $0.14\pm0.02\,\mathrm{Jy}\cdot\kms$, the latter of which is used in \citet{2023BenitezLlambay}.\ The former FAST value is listed in the FASHI \citep[FAST All-Sky \hi{}][]{2024Zhang} survey catalog (ID\_FASHI = 20230020346, Name = J125153.61+401759.1) along with a velocity width ($W_{50,FASHI}=32.54\pm1.92\kms$) that is commensurate with ours.\ While the GBT spectrum appears to have an excess of flux at lower heliocentric velocities (shaded and hatched region between $\mathrm{265\lesssim V_{helio} \lesssim 290\,\kms}$) relative to the FAST spectrum from \citetalias{2023Zhou}, the spectrum of CL9 from FASHI (J. Wang, priv. comm.) is similar in shape.\ Both the GBT and FASHI spectra of CL9 show signs of an asymmetric \hi{} profile with a steeper receding edge and a shallower approaching edge.\ The asymmetry in both the GBT and FASHI spectra suggest that it is real and not just a result of instrumental differences.\ While these new results do not preclude M94-CL9 from being classified as a RELHIC, it does bring it in some tension with the predicted properties of RELHICs which typically expect a narrow ($W_{50}\sim20\kms$), Gaussian-like \hi{} distribution.\ We remind the reader that our GBT detection is spatially unresolved, and  does not constrain the morphology of the \hi{} within the beam.\ Higher spatial resolution \hi{} data of M94-CL9 will be informative in this regard.\

\begin{acknowledgments}
We thank Ming Zhu, Ruilei Zhou, Alejandro Ben{\'\i}tez-Llambay, and Jing Wang for useful discussions.\ AK acknowledges support from the Natural Sciences and Engineering Research Council of Canada (NSERC), the University of Toronto Arts \& Science Postdoctoral Fellowship program, and the Dunlap Institute.\ KS acknowledges support from NSERC. The Green Bank Observatory is a facility of the National Science Foundation operated under cooperative agreement by Associated Universities, Inc.
\end{acknowledgments}

\vspace{5mm}
\facilities{GBT}

\software{\textsc{GBTIDL} \citep{2006GBTIDL},
          \textsc{Numpy} \citep{2020Harris}, 
          \textsc{Matplotlib} \citep{2007Hunter}
          }

\bibliography{gbtcl9}{}

\begin{thebibliography}{}
\expandafter\ifx\csname natexlab\endcsname\relax\def\natexlab#1{#1}\fi
\providecommand{\url}[1]{\href{#1}{#1}}
\providecommand{\dodoi}[1]{doi:~\href{http://doi.org/#1}{\nolinkurl{#1}}}
\providecommand{\doeprint}[1]{\href{http://ascl.net/#1}{\nolinkurl{http://ascl.net/#1}}}
\providecommand{\doarXiv}[1]{\href{https://arxiv.org/abs/#1}{\nolinkurl{https://arxiv.org/abs/#1}}}

\bibitem[{{Ben{\'\i}tez-Llambay} \& {Navarro}(2023)}]{2023BenitezLlambay}
{Ben{\'\i}tez-Llambay}, A., \& {Navarro}, J.~F. 2023, \apj, 956, 1, \dodoi{10.3847/1538-4357/acf767}

\bibitem[{{Ben{\'\i}tez-Llambay} {et~al.}(2017){Ben{\'\i}tez-Llambay}, {Navarro}, {Frenk}, {Sawala}, {Oman}, {Fattahi}, {Schaller}, {Schaye}, {Crain}, \& {Theuns}}]{2017BenitezLlambay}
{Ben{\'\i}tez-Llambay}, A., {Navarro}, J.~F., {Frenk}, C.~S., {et~al.} 2017, \mnras, 465, 3913, \dodoi{10.1093/mnras/stw2982}

\bibitem[{{Goddy} {et~al.}(2020){Goddy}, {Stark}, \& {Masters}}]{2020Goddy}
{Goddy}, J., {Stark}, D.~V., \& {Masters}, K.~L. 2020, Research Notes of the American Astronomical Society, 4, 3, \dodoi{10.3847/2515-5172/ab66bd}

\bibitem[{{Harris} {et~al.}(2020){Harris}, {Millman}, {van der Walt}, {Gommers}, {Virtanen}, {Cournapeau}, {Wieser}, {Taylor}, {Berg}, {Smith}, {Kern}, {Picus}, {Hoyer}, {van Kerkwijk}, {Brett}, {Haldane}, {del R{\'\i}o}, {Wiebe}, {Peterson}, {G{\'e}rard-Marchant}, {Sheppard}, {Reddy}, {Weckesser}, {Abbasi}, {Gohlke}, \& {Oliphant}}]{2020Harris}
{Harris}, C.~R., {Millman}, K.~J., {van der Walt}, S.~J., {et~al.} 2020, \nat, 585, 357, \dodoi{10.1038/s41586-020-2649-2}

\bibitem[{{Haynes} \& {Giovanelli}(1984)}]{1984Haynes}
{Haynes}, M.~P., \& {Giovanelli}, R. 1984, \aj, 89, 758, \dodoi{10.1086/113573}

\bibitem[{{Hunter}(2007)}]{2007Hunter}
{Hunter}, J.~D. 2007, Computing in Science and Engineering, 9, 90, \dodoi{10.1109/MCSE.2007.55}

\bibitem[{{Karunakaran} {et~al.}(2020){Karunakaran}, {Spekkens}, {Zaritsky}, {Donnerstein}, {Kadowaki}, \& {Dey}}]{2020bKarunakaran}
{Karunakaran}, A., {Spekkens}, K., {Zaritsky}, D., {et~al.} 2020, \apj, 902, 39, \dodoi{10.3847/1538-4357/abb464}

\bibitem[{{Marganian} {et~al.}(2006){Marganian}, {Garwood}, {Braatz}, {Radziwill}, \& {Maddalena}}]{2006GBTIDL}
{Marganian}, P., {Garwood}, R.~W., {Braatz}, J.~A., {Radziwill}, N.~M., \& {Maddalena}, R.~J. 2006, in Astronomical Society of the Pacific Conference Series, Vol. 351, Astronomical Data Analysis Software and Systems XV, ed. C.~{Gabriel}, C.~{Arviset}, D.~{Ponz}, \& S.~{Enrique}, 512

\bibitem[{{Zhang} {et~al.}(2024){Zhang}, {Zhu}, {Jiang}, {Cheng}, {Wang}, {Wang}, {Xu}, {Liu}, {Yu}, {Qian}, {Yu}, {Ai}, {Jing}, {Xu}, {Liu}, {Guan}, {Sun}, {Yang}, {Huang}, {Hao}, \& {FAST Collaboration}}]{2024Zhang}
{Zhang}, C.-P., {Zhu}, M., {Jiang}, P., {et~al.} 2024, Science China Physics, Mechanics, and Astronomy, 67, 219511, \dodoi{10.1007/s11433-023-2219-7}

\bibitem[{{Zhou} {et~al.}(2023){Zhou}, {Zhu}, {Yang}, {Yu}, {Yuan}, {Jiang}, \& {Xi}}]{2023Zhou}
{Zhou}, R., {Zhu}, M., {Yang}, Y., {et~al.} 2023, \apj, 952, 130, \dodoi{10.3847/1538-4357/acdcf5}

\end{thebibliography}
\bibliographystyle{aasjournal}

\end{document}